\documentclass{emulateapj}

\usepackage{color}

\def\ergcm2s{erg~s$^{-1}$}
\def\ergs{erg~s$^{-1}$}
\def\d5{$d_{5}$}
\def\d5^2{$d_{5}^{2}$}

\def\r15{r_{\rm 15}}

\def \deg{^\circ}

\def \hcm {\hbox {\ifmmode $ atom cm$^{-2}\else atom cm$^{-2}$\fi}}

\def\j11{\mbox{\object{IGR~J11014-6103}}}

\def \apjl {ApJL}
\def \apjs {ApJS}
\def \aap {A\&A}

\shorttitle{Gamma-rays from SS433}
\shortauthors{Bordas, Yang, Kalexhiu \& Aharonian.}

\begin{document}

\title{Detection of persistent gamma-ray emission toward SS433/W50}

\author{P. Bordas\altaffilmark{1}, R. Yang\altaffilmark{1}, E. Kafexhiu\altaffilmark{1} and F. Aharonian\altaffilmark{1, 2}}

\affil{$^{1}$~Max-planck-Institut f\"ur Kernphysik, Saupfercheckweg 1, 69117 Heidelberg, Germany \\
$^{2}$~School of Cosmic Physics, Dublin Institute for Advanced Studies, 31 Fitzwilliam Place, Dublin~2, Ireland
}
\email{pol.bordas@mpi-hd.mpg.de, ryang@mpi-hd.mpg.de}


\begin{abstract}

The microquasar SS433 features the most energetic jets known in our Galaxy. A large fraction of the jet kinetic power is delivered to the surrounding W50 nebula at the jet termination shock, from which high-energy emission and cosmic-ray production have been anticipated. Here we report on the detection of a persistent gamma-ray signal obtained with the {\it{Fermi}} Large Area Telescope {from an unidentified source which we tentatively associate, given its 99.9 \% confidence level position accuracy and the lack of any other high-energy emitter counterpart in the studied region, with SS433}. The obtained spectral energy distribution displays a distinct maximum at $\sim$~250 MeV {and extends only up to $\sim 800$~MeV}. We discuss the possibility that the observed gamma-ray emission is produced  through proton-proton collisions at the SS433/W50 interaction regions. If  the same mechanism is operating in other baryon-loaded microquasar jets, their collective contribution could represent a significant fraction of the total galactic cosmic-ray flux at GeV energies.

\end{abstract}



\section{Introduction}
\label{section:introduction}

Evidences of significant gamma-ray emission from a few accreting X-ray binary systems (XRBs) have been collected in the recent years, owing to the improved capabilities of ground- and space-based gamma-ray instruments (see, e.g., \citealp{Albert2007, Fermi2009, Ackermann2014}). In all cases, such detections have been associated to transient phenomena, with gamma-ray emission thought to be produced in the interaction of energetic outflows with the surrounding matter and radiation fields.
%
%
The kinetic energy carried by these outflows can increase significantly during such outbursts with respect to their quiescent state, with ejected kinetic luminosities reaching $L_{\rm kin}\sim 10^{37-38}$~erg~s$^{-1}$, sustained for periods ranging from a $\sim$few hours to some weeks ({see, e.g., \citealp{Mirabel1994, Atoyan1999, Revnivtsev2002, Heinz2002} and references therein}).

\noindent {SS433 is a well-known XRB system, and the first one in which highly energetic} bipolar {outflows} (jets) were discovered \citep{Abell1979, Fabian1979}. Identified 30 years ago as a bright  H$\alpha$ line-emitting star close to the Galactic plane \citep{Stephenson1977} and as a non-thermal radio \citep{Feldman1978} and X-ray source \citep{Marshall1978}, it has been since then object of intensive studies {as an unusually bright and powerful} microquasar \citep{MIrabel1999}.  
The system is composed of a compact object, most probably a 10--20~M$_{\odot}$ black hole \citep{Gies2002}, orbiting every 13.1 days {a most-probable $\sim$11~M$_{\odot}$ A3--7 I star in a circular orbit with radius $\sim$~56~R$_{\odot}$ \citep{Hillwig2004}}, and is distinguished by a continuous (non-transient) supercritical accretion regime onto the black hole, {with an accretion rate} $\sim 10^{-4}$~M$_{\odot}$~yr$^{-1}$ {\citep{vandenHeuvel1981}, with }\,M$_{\odot}$ and R$_{\odot}$ being the mass and radius of the sun, respectively. A supercritical accretion disk is therefore formed  \citep{Shakura1973}, 
which feeds two sub-relativistic jets carrying a kinetic energy power at a level of $L_{\rm kin}^{\rm jet}\sim$~10$^{39}$~erg~s$^{-1}$ \citep{Dubner1998}, ejected at a velocity $v_{\rm j} = 0.26 \, c$ \citep{Margon1989}, where $c$ is the speed of light. Both the jets and the disk are observed to precess every 162.4 days in cones of half opening angle $\theta_{\rm prec} \approx$ 21$^{\circ}$ with respect to the normal to the orbital plane, which in turn subtends an angle $i\approx$ 78$^{\circ}$ to our line of sight \citep{Eikenberry2001}.
In addition to the extremely large jet power, SS433 represents also one of only two cases in which the presence of baryonic material in the jets has been established (the other source being 4U 1630-47, \citealp{Diaz-Trigo2013}) through the detection of lines of highly ionized heavy elements \citep{Marshall2002, Migliari2002}.
Clouds of plasma containing both electrons, ions and protons propagate along ballistic trajectories whilst radiating intensively in radio, optical and X-rays. Eventually, the jets get decelerated through their interaction with the ambient medium through the formation of a termination shock. Most of the jet kinetic power is delivered at the shock interface to the surrounding W50 nebula, a large 2$^{\circ} \times 1^{\circ}$ size supernova remnant, SNR~G39.7$-$2.0 \citep{Green2009}, located at a distance of $\sim$~5 kpc \citep{Fabrika2004}. 
Radio to X-ray emission originated in the jet/nebula interaction sites has been proposed \citep{Safi-Harb1997, Dubner1998}, whereas gamma-ray radiation from these regions has also been predicted \citep{Heinz2002}, peaking at energies $\sim \Gamma_{\rm j}\,m_{\rm p}\,c^{2}$, where $\Gamma_{\rm j} = (1 - v_{\rm j}^{2}/c^{2})^{-1/2}$ is the bulk motion jet Lorentz factor. The detection of such a spectral feature would be a strong indication of cosmic-ray production in microquasar jets \citep{Heinz2002}. 
Very-high-energy gamma-ray emission has also been proposed \citep{Aharonian1998, Bosch-Ramon2005, Bordas2009} under the assumption that  particle acceleration at the jet termination shock extends to TeV energies. {No evidences of gamma-ray emission from SS433  has been reported so far} (see, e.g., \citealp{Aharonian2005, Hayashi2009, Saito2009}). 

\vspace{0.5cm}

\section{Fermi-LAT observations and analysis results}
\label{data}


{We have analysed archival $Fermi$-LAT observations, spanning about five years of data-taking (from MJD 54,682.6 to MJD 56,719.4), corresponding to observations of a $20\deg \times 20\deg$ region centred on SS433 (RA~=~$19^{\rm h} 11^{\rm m} 49^{\rm s}.57$, Dec.~=~$04^{\deg} 58' 57''.9$). This analysis has revealed the presence of a highly significant ($>  7 \sigma$) point-like source spatially coincident, within the 3$\sigma$ (99.9\%) confidence level position accuracy {of this newly discovered gamma-ray source, with the catalogued position of SS433}.} 
%
%
%
The data have been processed through the standard LAT analysis software
%
 %
 together with the latest version of the instrument response functions as well as the Galactic and isotropic diffuse models provided by the {\it Fermi}-LAT Collaboration\footnote{Software package v9r33p0, with the $\rm{P7rep_v15}$ version of post-launch instrument response functions (IRFs). Fermi Science Support Center (http://fermi.gsfc.nasa.gov/ssc)}. The sources present in the {3$^{\rm rd}$ $Fermi$-LAT catalogue} were included in the analysis, while allowing for a variation of the point-source parameters in the likelihood fit for sources inside the studied region. 
 %
 %
 A strong gamma-ray excess was apparent in the obtained residual maps. {We use {\it gtfindsrc} to derive the best fit position of this excess. The best fit position is  [RA$ = 287.65^{\circ}$, Dec$ = 4.72^{\circ}$], or about {$0.41^{\circ}$} away from SS433's nominal position, with a 1$\sigma$ error circle of 0.15$^{\circ}$ }. {The residuals showed in addition the presence of another uncatalogued point-like source candidate in the FoV, labeled ``ps2" hereafter, with best-fit position {[RA$ = 289.50^{\circ}$, Dec$ = 6.09^{\circ}$]}}.

 %
%
The inclusion of these two sources in the likelihood analysis, allowing again for a variation of their spectral parameters, provides a Test Significance value {TS = 57.6 above 200 MeV {(or about 7.3 $\sigma$)} for the source encompassing SS433, whereas a TS = 52.0 {(or $\sim 6.9 \sigma$)} is obtained for ``ps2" in our analysis}. {A $5^{\circ} \times 5^{\circ}$ TS map of the region of interest (ROI) is shown in Fig.~\ref{fig:TSmap}, together with the SS433 position and the 68\%, 95\% and 99.9\% confidence contours on the position of the gamma-ray source enclosing it.} 
%
%
{Given the Point Spread Function (PSF) of  the {\it Fermi}-LAT at energies }$E_{\gamma} \sim 300~\rm MeV$, above $\sim 1.5^{\circ}$, the $0.41^{\circ}$ difference of the centroid of the gamma-ray signal with respect to SS433 coordinates, and the error circle radius of the source position, of $\sim 0.15^{\circ}$, 
{the observed spacial coincidence prompts us for a tentative association of this new source with SS433}. 

\noindent A distinct feature of this detection is that the source shows a very unusual spectrum as compared to most of the {\it{Fermi}}-LAT catalog objects, with a spectral energy distribution (SED) displaying a maximum at $\sim$250 MeV, and extending only up to $\sim$800 MeV. 
%
%
{The gamma-ray data is best fitted with a \textit{logparabola} model, \mbox {$N_{0}(E/E_{\rm b})^{-(\alpha+ \beta \, log(E/E_{\rm b})}$}, with $N_{0} = 1.3 \pm 0.1 \times 10^{-7}$~cm$^{-2}$~s$^{-1}$~GeV$^{-1}$, $\alpha = 0.010 \pm 0.004$, $\beta = 1.9 \pm 0.1$, and $E_{\rm b} = 0.182 \pm 0.005$~GeV, yielding a Likelihood Test Ratio against a simple power-law of  LTR = -2log ($L_{\rm PL}$/$L_{\rm logpar}$)  = 13.2. A cutoff power-law also fits well the data, with a value LTR = 8.3 over a simple power-law fit.}

\noindent The bolometric gamma-ray flux above 200 MeV, { $ \Phi_{\gamma} = 2.4 \pm 0.3 \times 10^{-8}$~ph~cm$^{-2}$~s$^{-1}$}, does not show any significant variability in the $\sim$five-year data-set. We also searched for a possible periodic signature by folding the gamma-ray light-curve both on the orbital and precession periods of SS433. Due to the limited statistics, we divided all the available data in five equally spaced phase-bins in both cases. The flux has been derived by applying a likelihood fit above 200 MeV in each phase bin. The resulting light-curves, displayed in Fig.~\ref{fig:fold_lc}, are compatible with a constant flux.
\noindent To constrain the possible contamination of the signal by the bright, nearby $Fermi$-LAT pulsar 3FGL J1907.9+0602, we used the  {\it{Tempo2}} {pulsar timing package} \citep{hobbs06} and the latest ephemeris available for J1907.9+0602\footnote{Fermi Pulsar Timing (http://www.slac.stanford.edu/$\sim$kerrm/fermi\_pulsar\_timing)}, and performed the same likelihood analysis but using a restricted data-set including only photons falling in the {\it{off-pulse}} emission period of J1907.9+0602.
Using this $\sim 1/2$ exposure data-set, the source we associate to SS433 is detected at a TS value of 33 above $200~\rm MeV$. Both the total flux and the spectral properties are in complete agreement with respect to the analysis of the whole data-set, and we therefore make use of all available photons in the spectral analysis. The SED of the source associated to SS433 is shown in Fig.~\ref{fig:SED_combined_v2}.

\section{Discussion}

{The location of the newly discovered gamma-ray source, the lack of any other high-energy counterpart in the studied region, and the extreme energetics and known non-thermal activity of SS433, argue for an association of the two sources}. 

 Assuming a distance to SS433 of $d = 5\,d_{5}$~kpc, the derived flux {translates to} a gamma-ray luminosity $L_{\rm \gamma}\sim 7 \times 10^{34}$~\d5^2~\ergs, which is much lower than the bolometric {luminosity} of the system, $L_{\rm bol}\sim 10^{40}$~erg~s$^{-1}$ {\citep{Fabrika2004}}. The total power required to sustain it could be supplied close to the compact object by the supercritical accretion disk, or by the powerful jets both at the jet base or in the  region of interaction with the W50 nebula. Moreover, several emission mechanisms could contribute significantly to the gamma-ray flux. 

In a leptonic scenario, gamma-rays could be produced by Inverse Compton (IC) up-scattering off photons from the optical star and accretion disk or through relativistic Bremsstrahlung of electrons embedded in the cold proton/ion jet plasma or matter from the surrounding medium. 
%
%
{We consider in the following a simple power-law distribution for relativistic electrons with an exponential cutoff, \mbox {$N_{\rm e}\,\propto E_{\rm e}^{- \alpha_{\rm e}}$~exp\,($-E_{\rm e}/E_{\rm e, cut}$)}. More complex distributions for the emitting particles, e.g. a broken power-law, would require a proper knowledge of the acceleration conditions, possibly accounting for radiation or escape/adiabatic losses, which is beyond the scope of this study. Such a simple power-law distribution can fit the gamma-ray SED by assuming a hard electron spectrum in order to account for the low-energy data, at $E \lesssim 250$~MeV. Taking $\alpha_{\rm e}=1.5$, the best fit is obtained for a cutoff energy $E_{\rm e, cut} \approx$ 4.5 and 1.0~GeV in a IC and relativistic Bremsstrahlung scenario (see Fig.~\ref{fig:SED_combined_v2}), providing a $\chi^2$/d.o.f $= 4.99/3$ and $2.37/3$, respectively. Given the smooth {\it{logparabola}} best-fit to the gamma-ray data, we also tried sub/super exponential cutoffs in the electron distribution, with \mbox {   exp\,($-E_{\rm e}/E_{\rm e, cut}$)$^{\beta}$} and $\beta \in [0, 2] $, but no appreciable improvement in the fit is obtained}. 

{If the emitter is located inside or close to the binary-system, however}, periodic absorption features are expected to affect the observed gamma-ray flux every precession cycle (see, e.g., \citealp{Reynoso2008a}). For IC on the companion star photon field, an additional orbital modulation of the gamma-ray flux is expected given the high inclination of the system  ($i \approx 78\deg$). No signature of such modulation is observed neither in the precession nor the orbital phase-folded light-curves (see Fig.~\ref{fig:fold_lc}), although the low statistics available prevent from a definitive conclusion. 

If the emitter is instead located farther away along the jet, e.g. at the jet/medium interaction regions, IC on the cosmic microwave background would require a narrow electron distribution, approaching a delta function with $\gamma_{\rm e} \sim [3-5]\times 10^{5}$, to account for the observed gamma-ray flux.  {Relativistic Bremsstrahlung, on the other hand, could explain the observed emission by assuming target particle densities of $4 \times n_{\rm ion} \approx 4$~cm$^{-3}$, with $n_{\rm ion} \approx 1$~cm$^{-3}$ and a factor 4 from the Rankine-Hugoniot jump conditions in strong shocks, and a relatively high efficiency, of about $\sim 30$\%, in converting jet-kinetic energy to non-thermal energy . Such estimate for the non-thermal efficiency is nonetheless strongly dependent on the total energetic budget of the emitter, which is in any case highly uncertain, and therefore unsuitable to extract more robust conclusions.}

In a hadronic scenario, gamma-ray emission could be rendered through the decay of neutral pions produced in proton-proton collisions within the jet or at the jet termination site. The presence of a low-energy cutoff in the gamma-ray spectrum {could be} naturally explained in this scenario by the sharp decrease of the neutral pion production cross section close to the kinetic threshold, whereas the narrow gamma-ray peak {could} be  easily described by adopting a proton distribution which does not extend significantly beyond a few GeVs. 
{We have employed the latest parameterisation of the $pp \to \gamma$ cross sections given in \citet{Kafexhiu2014}, which is valid for proton energies from kinematic threshold up to PeV energies.}
%
%
Assuming for simplicity a power-law distribution \mbox {$N_{\rm p}\,\propto E_{\rm p}^{- \alpha_{\rm p}}$~exp\,($-E_{\rm p}/E_{\rm p, cut}$)} for relativistic protons, the SED can be well fit with best parameters $\alpha_{\rm p} \approx 1.9$ and $E_{\rm p, cut} \approx 4.2$~GeV.  
{We note that a similarly peaked gamma-ray spectrum, extending only up to $\sim1$~GeV, has been previously reported from the observations of solar flares with the $Fermi$-LAT \citep{Ackermann2014}, for which an hadronic has been favoured.} 

The jets of SS433 {could} naturally provide such a proton distribution, with related gamma-ray emission having been anticipated as a result of the jets interaction with the surrounding medium \citep{Heinz2002}.
A total energy $L_{\rm kin}^{\rm jet} \times t_{\rm SS433} \sim 6.3 \times 10^{\,50}$~ergs has been released by SS433 jets during the source life-time, $t_{\rm SS433} \sim 2 \times 10^{4}$~yrs. A fraction of this energy, $W_{\rm p}$, is expected to be thermalized when the cold jet particles cross the strong shock formed at the jet/medium interface, or transferred to a shock-accelerated particle population. The collision of protons behind the termination shock against  target nuclei from the shocked jet/medium shell {could lead} to gamma-ray {emission} through $\pi^{0}$ production and decay, with luminosities $L_{\rm \gamma}^{\rm \,pp} \sim W_{\rm p}/t_{\rm pp}$~erg~s$^{-1}$. 
For $\sim 1$~GeV protons the $pp$ interaction time-scale is $t_{\rm pp} \approx 1.2 \times 10^{15}$~s for target particle densities of 4~cm$^{-3}$. For the observed gamma-ray luminosity  $L_{\rm \gamma} \sim 7 \times 10^{34}$~erg~s$^{-1}$, setting  $L_{\rm \gamma} = L_{\rm \gamma}^{\rm pp}$ implies that a moderate $\sim$10\% fraction of the total accumulated energy {should} be contained in the injected protons and confined within the interaction regions.
Assuming that the confinement time is defined by diffusion, with $D(E) \approx 3 \times 10^{27}\,(E_{\rm p}/1~$GeV)$^{0.5}$~cm$^{2}$~s$^{-1}$,  and considering a length-scale for the interaction regions of $\Delta r \sim 10^{20}$~cm, the diffusion time-scale is $t_{\rm diff} \sim (\Delta r)^{2}/2\,D(E_{\rm p})\approx 5 \times 10^{4}$~yr, which is larger than the source age. 
{Note however that the estimations above assume that the source spectrum is primarily hadronic. An underlying contribution from Bremsstrahlung/IC to the observed flux cannot be ruled out, which would relax to some extent the energy requirements for the hadronic model.}

The thermalization of the jet proton's kinetic energy at the termination shock provides a Maxwell-Boltzmann distribution peaking at $k_{\rm B}\, T \sim (\Gamma_{\rm j}-1)\,m_{\rm p}c^{2} \sim 35$~MeV/nucleon for the sub-relativistic jets in SS433 (with $\Gamma_{\rm j} = 1.036$). {Even accounting for the interactions of protons in the high-energy tail of this thermal distribution}, such a temperature would be insufficient to fit the $\geq 250$~MeV data-points of the gamma-ray spectrum. An additional  population is therefore required, which could be supplied by accelerated protons  injected along the jet or at the jet termination shock. The acceleration mechanism should be on the other hand rather inefficient, {given that the proton spectrum should extend only up to a few GeVs}.  If acceleration takes place at the base of the jet, i.e., close to the compact object, these protons could emit gamma-rays through interactions with cold jet material or matter entrained from the disk wind. However, absorption of gamma-rays {through $\gamma N$ interactions} \citep{Reynoso2008a} should be apparent due to the periodic eclipses by the companion star and the precession of the accretion disk. Again, those absorption features are not observed in the phase-folded light-curves, within statistics (see Fig. \ref{fig:fold_lc}). A steady flux is expected instead if gamma-rays are produced by protons accelerated at the interaction shock of the jets with the surrounding W50 nebula.

The tentative association of the steady gamma-ray flux reported here with the microquasar SS433 may have only been possible given the extreme kinetic power of its jets, $L_{\rm kin}^{\rm \, SS433} \sim 10^{39}$~erg~s$^{-1}$.
%
%
For the same efficiency and {surrounding medium gas density}, any steady gamma-ray emission and cosmic-ray production from other less powerful microquasars will scale down in flux by a factor  {$\sim (L_{\rm \, kin}^{\rm  \, SS433}/L_{\rm kin}^{ \, \mu Q}) \, d_{5}^{-2} \sim$ 10$^{\, 2}$--10$^3$~$d_{5}^{-2}$, being undetectable by current gamma-ray instruments. 
However, the cumulative contribution from the microquasar population in our Galaxy, possibly much larger than the $\sim 20$ systems identified so far \citep{Fender2006}, {could account for a measurable fraction} of the galactic cosmic-ray flux at $\sim$~GeV energies \citep{Heinz2002} if jets from other systems are also baryon-loaded \citep{Diaz-Trigo2013}. 
{Furthermore, the jets of established microquasars have Lorentz factors in the range $\Gamma_{\rm j} \sim$~1--10. Collective effects would make the combined gamma-ray peak to appear correspondingly broader and shifted-up to $\sim$~few GeVs.
Interestingly, a pronounced spectral feature peaking at 1--4 GeV has been recently discovered in the inner regions of the Galaxy, extending for about 2--3~kpc from the Galactic Center \citep{Hooper-2013}.} 
{Galactic jets could therefore be contributing to this GeV feature, which has been otherwise interpreted in a  annihilating dark matter scenario \citep{Hooper2011}.} 
Further studies on the spatial and spectral distribution of a yet unresolved gamma-ray emitting microquasar population are required to test this hypothesis. 

{Finally, we have considered so far gamma-rays being produced either at the jet base or in their interaction with the surrounding W50 nebula. The observed fluxes could be otherwise originated by the W50 remnant itself. The Fermi-LAT has indeed detected about 20 gamma-ray emitting SNRs \citep{FermiLAT2015}. Such emission has been claimed to have an hadronic origin in at least two systems, W44 and IC443 \citep{Ackermann2013}. Both sources are rather bright, with gamma-ray luminosities above $\sim 10^{-10}$~erg~cm$^{-2}$~s$^{-1}$, and display spectral breaks at energies of $\sim$~few~$\times 100$~MeV, interpreted as the result of interactions with nearby molecular clouds. Despite their much broader spectra, extending well beyond the maximum $\sim 800$~MeV reported here (IC 443, in particular, is detected at TeV energies, \citealp{Albert2007}), this could be explained by a more efficient proton acceleration occurring in these systems with respect to W50. The factor $\sim 10$ times higher gamma-ray luminosity displayed by W44 and IC 443 could be on the other hand accounted for given their relative closer location, at 2.9 and 1.5~kpc, respectively, compared to the 5~kpc distance to W50. }


\acknowledgments{\small{
\noindent {Acknowledgments} The authors are grateful to Stanislav Kelner, Anton Prosekin, Andrea Tramacere, Yasunobu Uchiyama and Andrew M. Taylor for fruitful discussions during this work. {The authors are grateful as well to the anonymous referee, whose suggestions greatly helped to improve the manuscript}. The authors acknowledge the Max-Planck-Institut f\"ur Kernphysik financial support and excellent working conditions during this research.}}


\bibliographystyle{apj} 
\bibliography{bibliography}

\clearpage

\begin{figure*}[htb]
\centering
\includegraphics[width=120mm,angle=0]{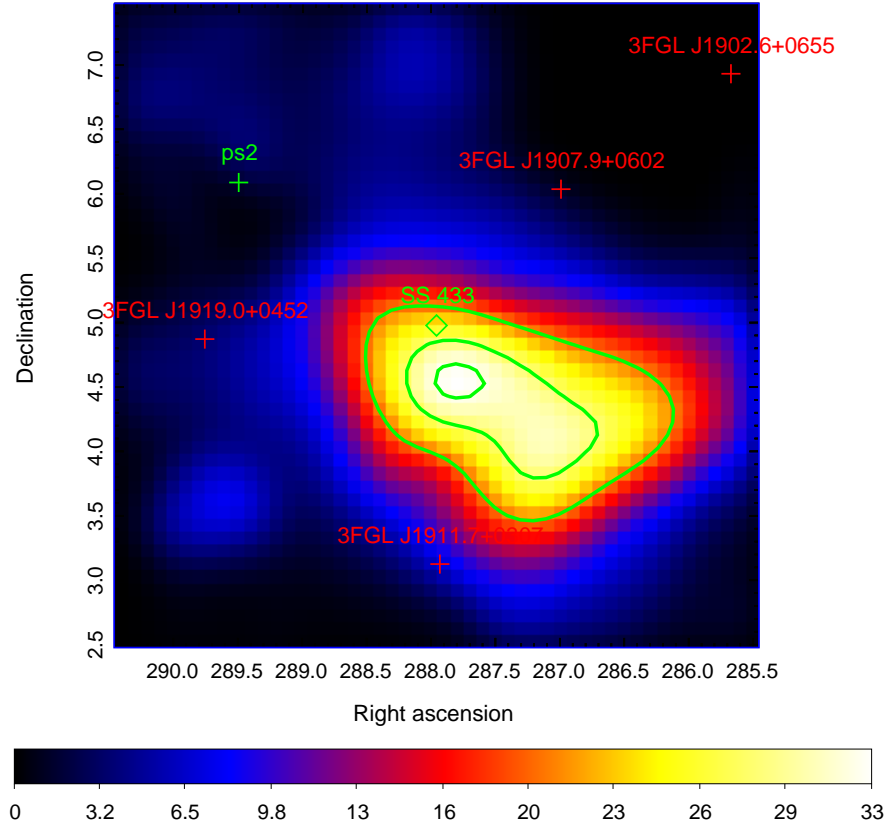}
\caption{Test Significance (TS) map above 300 MeV for the $5^{\circ}\times5^{\circ}$ ROI around the nominal position of SS433, with pixel size corresponding to $0.1^{\circ}\times0.1^{\circ}$. Sources included in the 3$^{\rm \, nd}$$Fermi$-LAT catalog located within the ROI, in particular the bright gamma-ray pulsar 3FGL J1907+0602, as well as the candidate source ``ps2" found in our analysis, have been excluded from the map for clarity. A green diamond indicates the position of SS433, whereas the green contours show the 68\%, 95\% and 99.9\% positional accuracy of its proposed $Fermi$-LAT counterpart, with a TS value of 57.6.} 
\label{fig:TSmap}
\end{figure*}

\begin{figure*}[htb] 
\centering 
\begin{minipage}[b]{0.45\linewidth}
\includegraphics[width=80mm,angle=0]{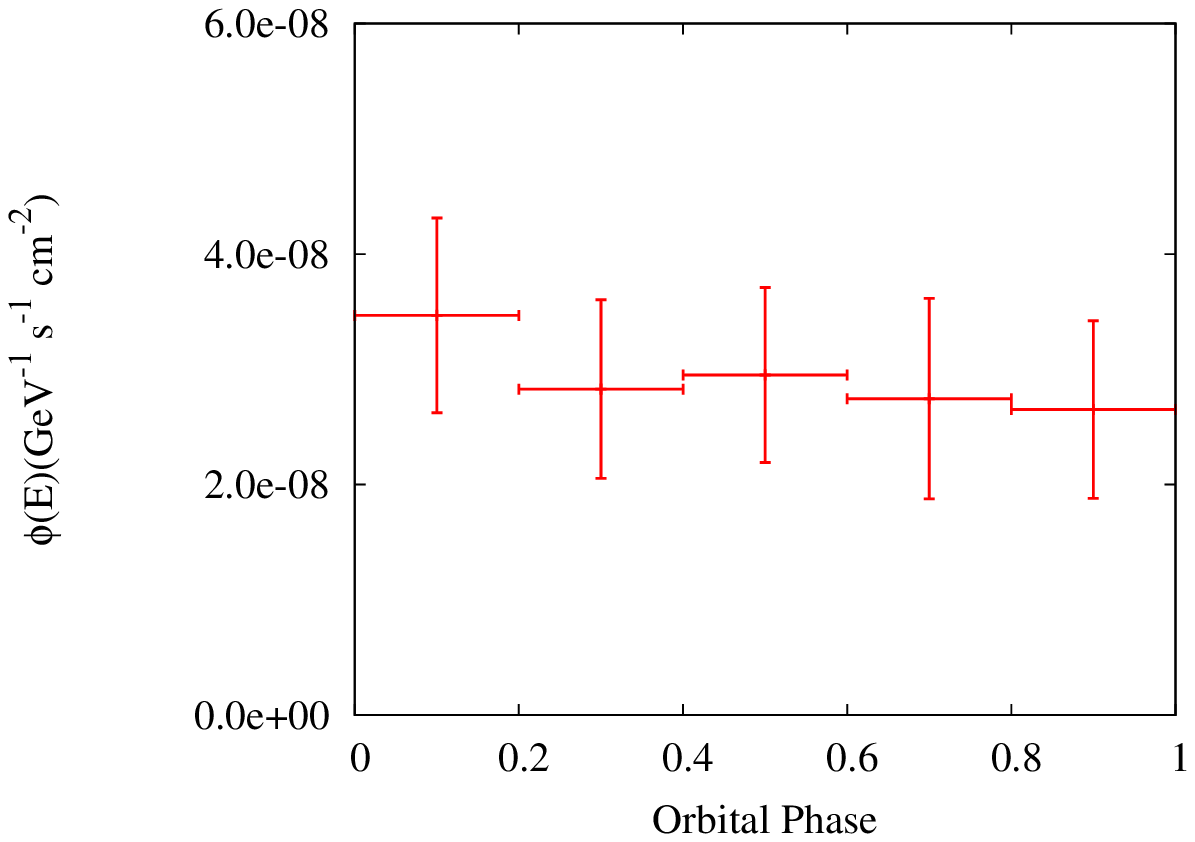}
\end{minipage} 
\quad \begin{minipage}[b]{0.45\linewidth}
\includegraphics[width=80mm,angle=0]{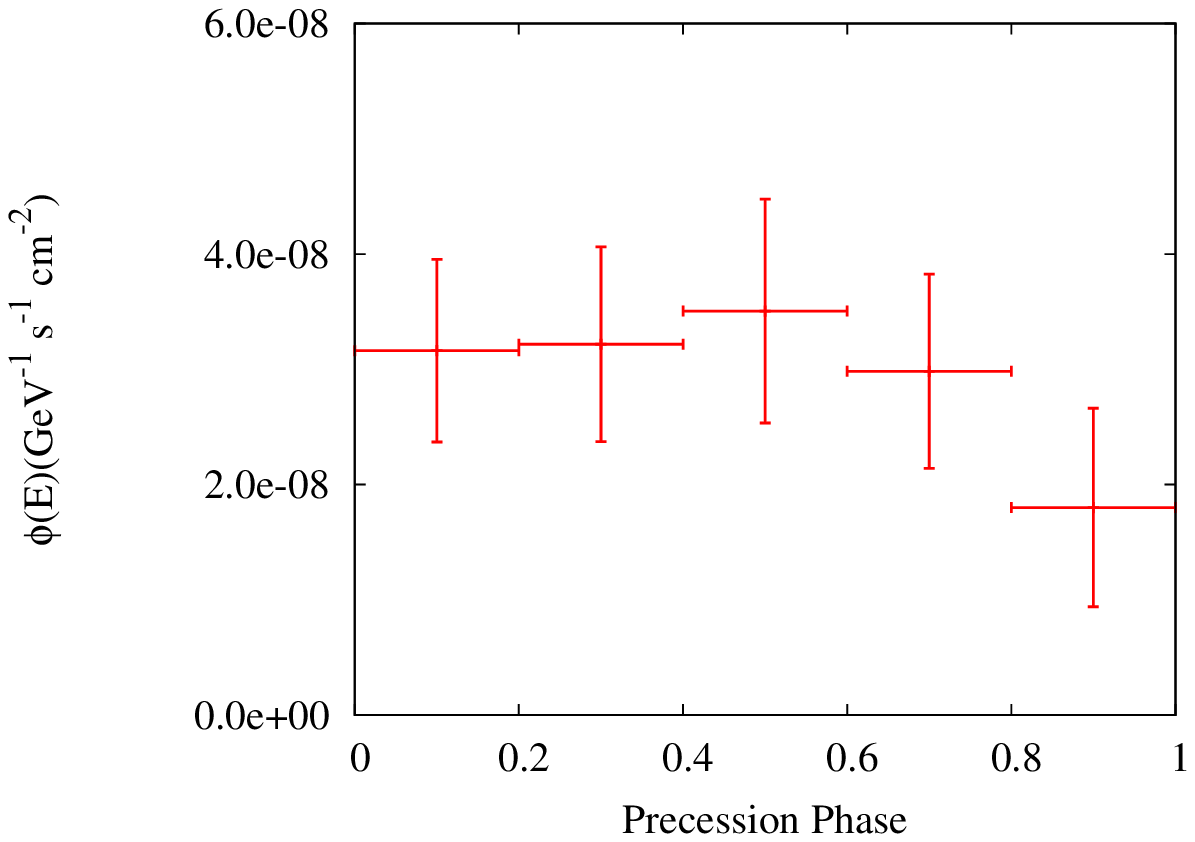}
\end{minipage} 
\caption{Phase-folded gamma-ray light-curve on the orbital and precession period of the system ($P = 13.15$ and $162.4$~days, respectively). Ephemeris by \citet{Gies2002} have been employed. No statistically significant signature of flux modulation is observed, with a constant fit to the data yielding $\chi^{2}$/d.o.f of 0.60/4 and 2.27/4 for the orbital and precession phase-folded light-curves, respectively.} 
\label{fig:fold_lc}
\end{figure*}

\begin{figure*}[htb]
\centering
\includegraphics[width=120mm,angle=0]{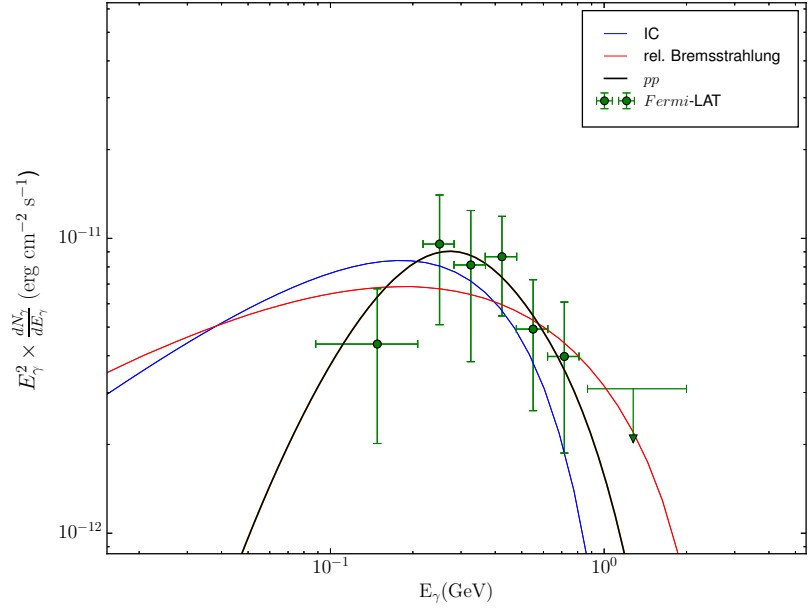}
\caption{Spectral energy distribution of the $Fermi$-LAT flux together with fits to the gamma-ray emission considering emission produced through IC (blue), relativistic  Bremsstrahlung (red) and proton-plasma interactions (black). 
%
%
The target photon field for IC is provided by the optical star and the accretion disk, ($T_{\star} = 8500$~K and $T_{\rm disk}= 40.000$~K, respectively \citep{Fabrika2004}). Target material for relativistic Bremsstrahlung comes from cold jet particles or material entrained from the surrounding medium. An electron power-law distribution with $\alpha_{\rm e} = 1.5$ and $E_{\rm cut} = 4.5 \, (1.0)$~GeV for IC (relativistic Bremsstrahlung) fits well the high-energy data points, but cannot reproduce the spectrum below $\sim 250$~MeV. For $pp$ interactions, a thermal proton distribution with $k_{\rm b}\,T \sim  35$~MeV is insufficient to match the high-energy data-points. A proton power-law extending up to a few GeVs is required to fit the spectrum.}
\label{fig:SED_combined_v2}
\end{figure*}

\end{document}